\documentclass[default,iicol]{sn-jnl}

\usepackage{accents}
\usepackage{siunitx}

\jyear{2021}%

\theoremstyle{thmstyleone}%
\theoremstyle{thmstyletwo}%

\theoremstyle{thmstylethree}%

\begin{document}

\title[Article Title]{Update of Muonium $1S-2S$ transition frequency}


\author*[1]{\fnm{Irene} \sur{Cortinovis} }\email{coirene@ethz.ch}
\author[1]{\fnm{Ben} \sur{Ohayon}} \email{bohayon@ethz.ch}
\author[1]{\fnm{Lucas} \sur{de Sousa Borges} }\email{lucasde@ethz.ch}
\author[1,2]{\fnm{Gianluca} \sur{Janka}} \email{gianluca.janka@psi.ch}
\author[1,3]{\fnm{Artem} \sur{Golovizin} }\email{artem.golovizin@gmail.com}
\author[1,3]{\fnm{Nikita} \sur{Zhadnov} }\email{nzhadnov@ethz.ch}
\author[1]{\fnm{Paolo} \sur{Crivelli}} \email{paolo.crivelli@cern.ch}

\affil[1]{\orgdiv{Institute for Particle Physics and Astrophysics}, \orgname{\textsc{eth}}, \orgaddress{\city{Zurich}, \postcode{8093}, \country{Switzerland}}}

\affil[2]{\orgdiv{Paul Scherrer Institute}, \orgname{PSI}, \orgaddress{\city{Villigen}, \postcode{5232}, \country{Switzerland}}}

\affil[3]{\orgdiv{P.N. Lebedev Physical Institute}, \orgaddress{\city{Moscow}, \postcode{119991}, \country{Russia}}}

\abstract{
We present an updated value of the Muonium 1S-2S transition frequency, highlighting contributions from different QED corrections as well as the large uncertainty in the Dirac contribution, stemming from the uncertainty of the electron to muon mass ratio. Improving the measurement of this spectral line would allow to extract a more accurate determination of fundamental constants, such as the electron to muon mass ratio or, combined with the Muonium hyperfine splitting, an independent value of the Rydberg constant. Furthermore, we report on the current status of the Mu-MASS experiment, which aims at measuring the Muonium 1S-2S transition frequency at a \SI{10}{\kilo\hertz} uncertainty level. 
}

\maketitle

\section{Introduction}\label{sec1}
Muonium (M) is an exotic bound state of an antimuon ($\mu^+$) and an electron ($e^-$). Being a purely leptonic system devoid of internal structure and nuclear finite size effects, Muonium lays an ideal playground to test quantum electrodynamics (QED) \cite{2005_Karshenboim}. In the scope of this proceeding, we will focus on the Muonium $1S-2S$ spectral line $\nu_{1S-2S}$.
Compared to Positronium (Ps), its relatively long lifetime (\SI{2.2}{\micro\second}) and larger mass make Muonium an attractive candidate for spectroscopy measurements. Owing to the longer lifetime of Muonium, the $1S-2S$ transition is more narrow (\SI{145}{\kilo\hertz}) than in Ps (\SI{1.26}{\mega\hertz}). Additionally, experimenting with heavier atoms is easier since at a given temperature they move slower.

Currently, the best measurement of the Muonium $1S-2S$ transition is 2455528941.0(9.8) MHz \cite{2000_Meyer}, in good agreement with the QED prediction of 2455528935.4(1.4) MHz \cite{1996-Pachucki}.

Advancing the experimental precision of this transition has multiple motivations. For instance, it will lead to the most precise value of the electron to muon mass ratio. Alternatively, together with the ongoing efforts for improving the hyperfine splitting \cite{2020_MUSEUM}, it will give the opportunity to test bound state QED, or the possibility to extract the Rydberg constant independently of nuclear and finite-size effects. Taking the Rydberg constant from hydrogen spectroscopy, Muonium spectroscopy offers a possibility to independently determine the muon g-2 with sufficient accuracy to contribute to the understanding of the current discrepancy \cite{delaunay2021towards}. Moreover, this measurement could reach interesting sensitivity to possible New Physics scenarios, such as Lorentz- and CPT-violations in the context of the Standard Model Extension (SME) \cite{2014_Vargas}, or new forces mediated by light bosons coupled to muons and electrons \cite{2019_Dark}, as well as provide a stringent test of lepton universality, by probing the muon to electron charge ratio below the current ppb level limit \cite{2000_Meyer}.

The Mu-MASS collaboration aims to measure the $1S-2S$ transition in Muonium with a final uncertainty of \SI{10}{\kilo\hertz} \cite{2018_Crivelli}, providing a 1000-fold improvement on the state of the art. 
The current best measurement is limited by the MHz level uncertainties brought by the pulsed laser which drives the $1S-2S$ transition, mainly due to the laser chirp and to the residual first linear Doppler shift. Additionally, the limited interaction time due to the laser pulse results in an intrinsic linewidth broadening. By using a continuous wave (CW) laser, the measurement is free from these limitations at the cost of having a lower excitation probability. However, the progress on the UV CW laser technology \cite{2021_Burkley}, together with the unique flux of low energy muons available at the Low Energy Muon (LEM) beamline at PSI \cite{2008_Prokscha}, and new methods for efficient and slow Muonium formation in vacuum \cite{2012-Mesoporous, Antognini:2022ysl}, open up the possibility for large improvements on the measured transition.

Such a potential leap in the experimental accuracy calls for an update on the theoretical value of the transition. Since the latest estimation, considerable advancements in the QED calculations were made \cite{karshenboim2019lamb,yerokhin2019theory,adkins2023recoil,2022-Eides, karshenboim2022complete}. For the purpose of determining the electron to muon mass ratio, it is convenient to decouple the uncertainty depending on the electron to muon mass ratio (currently dominating), from the smaller contribution depending on the QED calculations, which latest estimation is \SI{20}{\kilo\hertz} \cite{2000_Meyer}. In this way, one can conveniently compute the uncertainty in the value of the $1S-2S$ transition for any given assumption of electron to muon mass ratio uncertainty. Additionally, if the electron to muon mass ratio could be determined at the part per billion (ppb) level by an improvement in the Muonium HFS measurement \cite{strasser2019new}, and the experimental $1S-2S$ uncertainty would reach the kHz level, one will be able to test the QED corrections.

\section{Calculation of Muonium $1S-2S$ transition frequency}\label{sec2}
The energy levels for Muonium in a given principal quantum number $n$ satisfy :
\begin{equation}\label{eq:En}
E_n = - \frac{R_{\infty} c}{n^2 (1+m_e/m_{\mu})} (1 + \mathcal{F}),
\end{equation}
where $\mathcal{F} \ll 1$ takes into account higher order corrections such as recoil and QED \cite{2018-CODATA}.

The largest contribution to the M $1S-2S$ transition energy is given by the Dirac eigenvalue for an electron bound to a muon. By denoting $m_r$ the reduced mass of the electron-muon system, and $M$ the total mass of the atom $m_e + m_{\mu}$, the Dirac contribution $E_{\mathrm{Dirac}}$ \cite{2018-CODATA}  is:
\begin{align}\label{eq:dirac}
E_{\mathrm{Dirac}} &=   M c^2 + (f(n,J,\alpha)-1) m_r c^2 - \nonumber\\
&(f(n,J,\alpha)-1)^2 \frac{m_r^2 c^2}{2 M} , 
\end{align}
where $f(n, J, \alpha)=\left[1+\frac{(Z \alpha)^2}{(n-\delta(J, \alpha))^2}\right]^{-1 / 2}$, and $ \delta(J, \alpha) = J + \frac{1}{2} - \sqrt{(J + \frac{1}{2})^2 - (Z \alpha)^2} $.

Using the current best QED-independent experimental value for the ratio of the masses, namely $ \frac{m_{\mu}}{m_e} = 206.768 277(24)$ (120 ppb) from the measurement of the muon
magnetic moment determined by the Rabi method \cite{liu1999high}, the calculation of $E_{\mathrm{Dirac}}$ for Muonium yields 2455535991.3(1.4) MHz. The uncertainty is almost entirely due to our knowledge of the ratio of masses, and dominates the total uncertainty of $\nu_{1S-2S}$. When the electron to muon mass ratio will be measured experimentally with a higher accuracy, the uncertainty from the Dirac contribution will accordingly decrease. Alternatively, from a better experimental uncertainty of the M $1S-2S$ transition, one can extract the electron to muon mass ratio with higher precision.
To quantify this, one can use Equation \ref{eq:En} and obtain the relation between the relative uncertainties of the electron to muon mass ratio and of $\nu_{1S-2S}$. As a first approximation we use that $\mathcal{F} \ll 1$ to obtain 
\begin{equation}
    \nu_{1S-2S} \approx \frac{3}{4} \frac{R_{\infty} c}{1 + m_e/m_{\mu}}.
\end{equation} 
Secondly, we express the error on  $\nu_{1S-2S}$ neglecting the smaller contributions related to the Rydberg constant uncertainty:
\begin{align*}
\sigma_{\nu_{1S-2S}} &\approx   \frac{3}{4} \frac{R_{\infty} c}{(1 + m_e/m_{\mu})^2} \cdot \sigma_{m_e/m_{\mu}}\\
&\approx \nu_{1S-2S} \cdot \sigma_{m_e/m_{\mu}} , 
\end{align*}
where in the last step we assumed $\frac{m_e}{m_{\mu}} \ll 1$. Rearranging and dividing both sides by $\frac{m_e}{m_{\mu}}$, one obtains:
\begin{equation}\label{eq:uncert}
    \frac{\sigma_{m_e/m_{\mu}}}{m_e/m_{\mu}} \approx \frac{\sigma_{\nu_{1S-2S}}}{\nu_{1S-2S}} \cdot \frac{m_{\mu}}{m_e}.
\end{equation} 
Equation \ref{eq:uncert} can be used with the value of an experimentally measured $\nu_{1S-2S}$ to determine the relative uncertainty of the electron to muon mass ratio obtained from the measurement itself. For example, when a \SI{10}{\kilo\hertz} uncertainty will be reached for $\nu_{1S-2S}$, the electron to muon mass ratio will be determined to the level of 1 ppb.

Additionally to the Dirac energy, there are numerous other smaller contributions to the final value of $\nu_{1S-2S}$, summarised in Table \ref{tab:calculations_summary}. Their expressions are described in detail for the case of the M Lamb shift in \cite{2022_Janka}.
In first approximation these contributions are 7 times larger than for the Lamb shift due to their $\frac{1}{n^3}$ dependency. Another difference is that for the $1S-2S$ transition the Barker-Glover and off-diagonal hyperfine-structure contributions are zero.
Furthermore, an updated calculation for $E_\mathrm{rec,R2}$, namely the expansion in mass ratio of the pure recoil term of order $(Z\alpha)^6$ from \cite{adkins2023recoil, pachucki1997recoil}, removes the uncertainty given from the fact that the formula used in \cite{2022_Janka} was incomplete.
Finally, we include higher orders (i.e. the term with the A50 coefficient, calculated from the the $Z \alpha$ expansion of one-loop self energy \cite{2018-CODATA, 2019_Pachucki}) in the muon self energy $E_\mathrm{SEN}$ \cite{eides2007theory}. 
Overall, the updated value for the QED contributions to $\nu_{1S-2S}$ adds up to -7056.062(6) MHz, where the correction that dominates the uncertainty is the radiative recoil $E_\mathrm{RR}$, due to the uncomputed coefficient of the term of the order $\alpha (Z \alpha)^5 (Z \alpha) \ln{(Z \alpha)}^{-2}$ \cite{2018-CODATA}.

\begin{table}[h]

  \centering
    \begin{tabular}{l|lr}
    \hline \noalign{\vspace{-1pt}} \hline
        \noalign{\smallskip}
                Contr. & Largest Order & Muonium \\
                & & (MHz) \\
        \noalign{\smallskip}
         \hline
        \noalign{\smallskip}
$E_\text{Dirac}$  & $(Z\alpha)^2$  & $2455535991.3(1.4)$ \\\\
$E_\text{SE}$  & $\alpha~(Z\alpha)^4$  & $-7222.771$ \\\\
$E_\text{VP}$  & $\alpha~(Z\alpha)^4$  & $185.565$ \\
$E_{\text{VP}\mu+\text{had}}$ & $\alpha~(Z\alpha)^4(m_e/m_\mu)^2$  & $0.007$ \\\\
$E_\text{2ph}$ & $\alpha^2(Z\alpha)^4$  & $-0.627(1)$ \\
$E_\text{3ph}$ & $\alpha^3(Z\alpha)^4$   & $-0.001$ \\\\
$E_\text{rec,S}$  & $ ~ ~ ~ (Z\alpha)^5~(m_e/m_n)$  & $-18.104$ \\
$E_\text{rec,R}$  & $ ~ ~ ~ (Z\alpha)^6~  (m_e/m_n)$  & $0.056$\\
$E_\text{rec,R2}$ & $ ~ ~ ~ (Z\alpha)^6~ (m_e/m_n)^2$  & 0.005 \\
$E_\text{RR}$     & $\alpha~(Z\alpha)^5~ (m_e/m_n)$   & $0.095(6)$ \\
$E_\text{RR2}$     & $\alpha~(Z\alpha)^5~ (m_e/m_n)^2$   & $-0.001$ \\
\\

$E_\text{SEN}$ & $Z^2\alpha(Z\alpha)^4(m_e/m_n)^{2}$ &  $-0.286$\\ 

    \noalign{\smallskip}
    \hline \noalign{\smallskip} \hline
    \noalign{\vspace{2pt}}
    
Sum &  & $2455528935.2(1.4)$ \\
QED only  &  & $-7056.062(6)$ \\

    \noalign{\smallskip}
    \hline \noalign{\smallskip} \hline
    \noalign{\vspace{2pt}}

\end{tabular}
\caption[]{Summary of the calculated contributions to the Muonium $1S-2S$ transition. Uncertainties smaller than \SI{0.5}{\kilo\hertz} are not tabulated. The notation refers to the definitions in \cite{2022_Janka}.}
\label{tab:calculations_summary}
\end{table}

\section{Experimental methods}\label{sec3}
\begin{figure}[h]
\centering
\includegraphics[width=1\columnwidth,trim={0 0 0 0},clip]{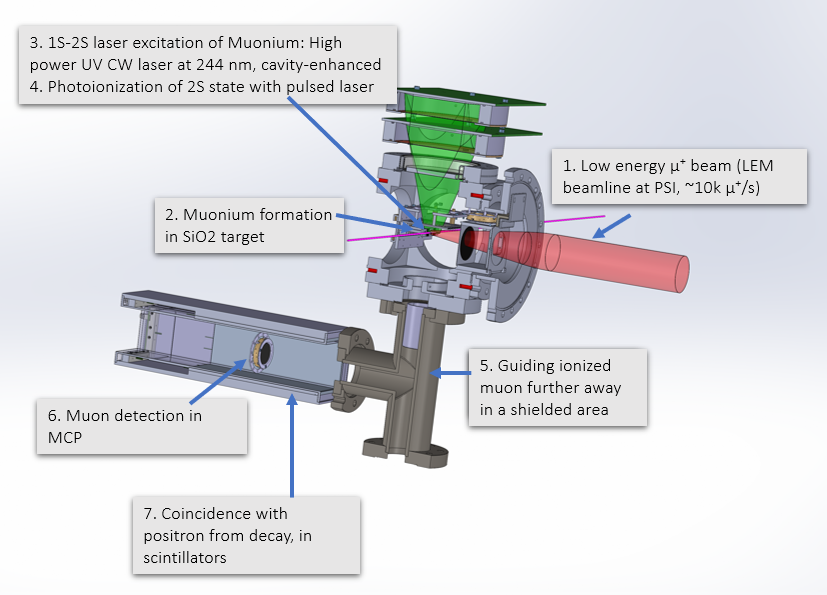}
    \caption{Schematic view of the Mu-MASS $1S-2S$ setup.}
\label{fig:setup}
\end{figure}
The Mu-MASS experiment runs at the LEM beamline of PSI, which provides a pure, low energy (selectable between 1 and 30 keV) $\mu^+$ beam \cite{2008_Prokscha}. A sketch of the experimental setup is shown in Figure \ref{fig:setup}. Approximately \SI{3}{\kilo\hertz} $\mu^+$ are tagged event by event, and focused to a $\sigma \approx$ \SI{4}{\milli\meter} wide beam, impinging onto a mesoporous thin SiO\textsubscript{2} film target. Here, thermalized Muonium can be formed and emitted into vacuum \cite{2012-Mesoporous}.
When Muonium traverses the \SI{244}{\nano\meter} CW laser, it can be excited to the $2S$ state via two-photon excitation, and further be photoionized by a \SI{355}{\nano\meter} pulse. The photoionized $\mu^+$ is then electrostatically guided to a microchannel plate (MCP) detector. Furthermore, to suppress background, four scintillator detectors surround the MCP area, to detect the positron from the $\mu^+$ decay. Therefore, the experimental signature of a $2S$ excited $\mu^+$ consists of a triple coincidence well defined in time, namely: a count in the tagging detector (which provides the initial time), a count in the photoionized  $\mu^+$ MCP detector after the expected time of flight from the \SI{355}{\nano\meter} laser pulse, and a count in one of the scintillators within few microseconds. A lineshape is obtained by measuring the rate of the detected $2S$ M candidates while varying the CW laser frequency referenced to a GPS disciplined frequency comb, and from this the Muonium $\nu_{1S-2S}$ resonance transition frequency is extracted.

As mentioned above, the laser system of the Mu-MASS project consists of two main parts: CW laser at the wavelength \SI{244}{\nano\meter} for two-photon excitation of 1S-2S transition and pulsed laser at the wavelength \SI{355}{\nano\meter} for photoionization of Muonium in 2S state. The first laser, which is a custom-designed commercial system with a home-built second harmonic generation cavity, can provide more than 1.5 W of UV output power \cite{2021_Burkley}, which can be enhanced by a factor of more than 30 inside the research vacuum chamber with the help of a Fabry-Perot cavity. Due to very low excitation probability, the laser system is required to work stably for periods of up to 1 week to collect the proper amount of data. That is not possible with constant operation at full laser power due to mirror degradation from a high-power UV radiation \cite{2021_Burkley}. To control the laser power, an AOM is used, which allows us to turn on the maximum intensity of 488\,nm light, and consequently  UV radiation in the enhancement cavity only for a short time after we get a signal from the tagging detector. Between these sharp increases, the power is kept at a few milliwatts to maintain the enhancement cavity locked to the laser wavelength. That allows running the \SI{244}{\nano\meter} laser at a low enough average power to prevent mirrors from fast degradation. Another feature of this method is the ability to turn off the laser radiation entirely for the time window when we expect to detect the photoionized muon. That allowed us to decrease laser-induced background noise. Power changes described above happen at times not exceeding $2 \mu s$ and do not violate the locking of the cavity to the laser. After excitation, the \SI{355}{\nano\meter} laser emits a pulse with an energy of more than 1 mJ. With several passes through the chamber, it provides almost hundred-percent photoionization. 

The described method was used during measurements on the LEM beamline at PSI. Even though it does not completely avoid mirror degradation, it can significantly reduce its rate. We were able to to maintain the radiation power in the enhancement cavity from 25 to 15 W for five days, conditioning the mirrors with oxygen approximately 2-4 times a day.

What makes this measurement extremely challenging is the low excitation probability of the CW laser. The Muonium $2S$ signal rate in the Mu-MASS setup depends quadratically on the \SI{244}{\nano\meter} laser power and scales linearly with the initial $\mu^+$ rate. With \SI{25}{\watt} of continuous laser power on resonance and the foreseen improvements in the muon tagging system we expect to have around 1 event per hour. 
It is therefore key to keep the background (coming from accidental counts in the detectors, muon-induced or laser-induced) as low as possible. The demonstrated background rates are consistent with less than 1 background event per day.

\section{Conclusions}\label{sec4}
We presented an updated value of the $1S-2S$ transition in Muonium, separating the smaller QED contributions from the Dirac energy. Concerning the QED part, the final result is consistent with the literature and shows an improvement of more than a factor 2 on the uncertainty latest estimations \cite{2000_Meyer,delaunay2021towards}. 
We also outlined the status of the Mu-MASS experiment, which aims to measure the $1S-2S$ transition with a CW laser, at the LEM beamline at PSI. The experiment is extremely challenging, being the signal rate of the order of a few events per day, due to the highly suppressed excitation rate and the limited initial muon statistics. For this reason, the background levels has to be kept as low as possible. Tests at PSI showed that we can achieve less than 1 background event per day demonstrating the feasibility of the experiment. In the near future, improvements on the available statistics of muons are expected from the developments at PSI on the LEM beamline (such as upgrading the surface muon beamline \cite{zhou2022simulation} and an improved tagging detector with a thin carbon foil), or from the MuCool project \cite{antognini2021mucool}, and possibly by additional orders of magnitude from the HIMB upgrade \cite{2021_HiMB2}.

\section*{Acknowledgements}\label{sec5}
This work is based on experiments performed at the Swiss Muon Source S$\mu$S, Paul Scherrer Institute, Villigen, Switzerland. This work is supported by the ERC consolidator grant 818053-Mu-MASS (P.C.) and the Swiss National Science Foundation under grant 197346 (PC). B.O. acknowledges support from the European Union’s Horizon 2020 research and innovation program under the Marie Skłodowska-Curie grant agreement No. 101019414. 

\section*{Authors contribution}
All authors contributed to the study and the experiment. Material preparation, data collection
and analysis were performed by all authors. The first draft of the manuscript was written by IC, BO and NZ and all authors commented on previous versions of the manuscript. All authors read and approved the final manuscript.

\section*{Data availability statement}
The datasets generated during and/or analysed during the current study are available from the corresponding author on reasonable request.

\bibliographystyle{unsrt}
\bibliography{sn-article}

\begin{thebibliography}{10}

\bibitem{2005_Karshenboim}
Savely~G. Karshenboim.
\newblock {Precision physics of simple atoms: QED tests, nuclear structure and
  fundamental constants}.
\newblock {\em Phys. Rept.}, 422:1--63, 2005.

\bibitem{2000_Meyer}
V.~Meyer et~al.
\newblock {Measurement of the 1s - 2s energy interval in muonium}.
\newblock {\em Phys. Rev. Lett.}, 84:1136, 2000.

\bibitem{1996-Pachucki}
K.~Pachucki, D.~Leibfired, M.~Weitz, A.~Huber, W.~Konig, and T.~W. Hansch.
\newblock {Theory of the energy levels and precise two photon spectroscopy of
  atomic hydrogen and deuterium}.
\newblock {\em J. Phys. B}, 29:177--195, 1996.

\bibitem{2020_MUSEUM}
S.~Nishimura et~al.
\newblock {Rabi-oscillation spectroscopy of the hyperfine structure of muonium
  atoms}.
\newblock {\em Phys. Rev. A}, 104(2):L020801, 2021.

\bibitem{delaunay2021towards}
C\'edric Delaunay, Ben Ohayon, and Yotam Soreq.
\newblock {Towards an Independent Determination of Muon g-2 from Muonium
  Spectroscopy}.
\newblock {\em Phys. Rev. Lett.}, 127(25):251801, 2021.

\bibitem{2014_Vargas}
Andr\'e~H. Gomes, Alan Kosteleck\'y, and Arnaldo~J. Vargas.
\newblock {Laboratory tests of Lorentz and CPT symmetry with muons}.
\newblock {\em Phys. Rev. D}, 90(7):076009, 2014.

\bibitem{2019_Dark}
Claudia Frugiuele, Jes\'us P\'erez-R\'\i{}os, and Clara Peset.
\newblock {Current and future perspectives of positronium and muonium
  spectroscopy as dark sectors probe}.
\newblock {\em Phys. Rev. D}, 100(1):015010, 2019.

\bibitem{2018_Crivelli}
P.~Crivelli.
\newblock {The Mu-MASS (MuoniuM lAser SpectroScopy) experiment}.
\newblock {\em Hyperfine Interact.}, 239(1):49, 2018.

\bibitem{2021_Burkley}
Zakary Burkley, Lucas de~Sousa Borges, Ben Ohayon, Artem Golovozin, Jesse
  Zhang, and Paolo Crivelli.
\newblock {Stable high power deep-uv enhancement cavity in ultra-high vacuum
  with fluoride coatings}.
\newblock {\em Opt. Express}, 29(17):27450--27459, 2021.

\bibitem{2008_Prokscha}
T.~Prokscha, E.~Morenzoni, K.~Deiters, F.~Foroughi, D.~George, R.~Kobler,
  A.~Suter, and V.~Vrankovic.
\newblock {The new muE4 beam at PSI: A hybrid-type large acceptance channel for
  the generation of a high intensity surface-muon beam}.
\newblock {\em Nucl. Instrum. Meth. A}, 595:317--331, 2008.

\bibitem{2012-Mesoporous}
A.~Antognini et~al.
\newblock {Muonium emission into vacuum from mesoporous thin films at cryogenic
  temperatures}.
\newblock {\em Phys. Rev. Lett.}, 108:143401, 2012.

\bibitem{Antognini:2022ysl}
A.~Antognini et~al.
\newblock {Room-temperature emission of muonium from aerogel and zeolite
  targets}.
\newblock {\em Phys. Rev. A}, 106(5):052809, 2022.

\bibitem{karshenboim2019lamb}
Savely~G. Karshenboim, Akira Ozawa, Valery~A. Shelyuto, Robert Szafron, and
  Vladimir~G. Ivanov.
\newblock {The Lamb shift of the 1$s$ state in hydrogen: Two-loop and
  three-loop contributions}.
\newblock {\em Phys. Lett. B}, 795:432--437, 2019.

\bibitem{yerokhin2019theory}
Vladimir~A. Yerokhin, Krzysztof Pachucki, and Vojtech Patkos.
\newblock {Theory of the Lamb Shift in Hydrogen and Light Hydrogen-Like Ions}.
\newblock {\em Annalen Phys.}, 531(5):1800324, 2019.

\bibitem{adkins2023recoil}
Gregory~S. Adkins, Jonathan Gomprecht, Yanxi Li, and Evan Shinn.
\newblock {Recoil Corrections to the Energy Levels of Hydrogenic Atoms}.
\newblock {\em Phys. Rev. Lett.}, 130(2):023004, 2023.

\bibitem{2022-Eides}
Michael~I. Eides and Valery~A. Shelyuto.
\newblock {Three-loop corrections to the Lamb shift in muonium and
  positronium}.
\newblock {\em Phys. Rev. A}, 105(1):012803, 2022.

\bibitem{karshenboim2022complete}
S.~G. Karshenboim, A.~Ozawa, V.~A. Shelyuto, E.~Yu. Korzinin, R.~Szafron, and
  V.~G. Ivanov.
\newblock {The Complete \ensuremath{\alpha}$^{8}$m Contributions to the 1s Lamb
  Shift in Hydrogen}.
\newblock {\em Phys. Part. Nucl.}, 53(4):773--786, 2022.

\bibitem{strasser2019new}
P.~Strasser et~al.
\newblock {New precise measurements of muonium hyperfine structure at J-PARC
  MUSE}.
\newblock {\em EPJ Web Conf.}, 198:00003, 2019.

\bibitem{2018-CODATA}
Eite Tiesinga, Peter~J. Mohr, David~B. Newell, and Barry~N. Taylor.
\newblock {CODATA recommended values of the fundamental physical constants:
  2018*}.
\newblock {\em Rev. Mod. Phys.}, 93(2):025010, 2021.

\bibitem{liu1999high}
Weiwen Liu et~al.
\newblock {High precision measurements of the ground state hyperfine structure
  interval of muonium and of the muon magnetic moment}.
\newblock {\em Phys. Rev. Lett.}, 82:711--714, 1999.

\bibitem{2022_Janka}
Gianluca Janka, Ben Ohayon, and Paolo Crivelli.
\newblock {Muonium Lamb shift: theory update and experimental prospects}.
\newblock {\em EPJ Web Conf.}, 262:01001, 2022.

\bibitem{pachucki1997recoil}
Krzysztof Pachucki.
\newblock {Recoil Effects in Positronium Energy Levels to Order alpha6}.
\newblock {\em Phys. Rev. Lett.}, 79:4120--4123, 1997.

\bibitem{2019_Pachucki}
Vladimir~A. Yerokhin, Krzysztof Pachucki, and Vojtech Patkos.
\newblock {Theory of the Lamb Shift in Hydrogen and Light Hydrogen-Like Ions}.
\newblock {\em Annalen Phys.}, 531(5):1800324, 2019.

\bibitem{eides2007theory}
Michael~I. Eides, Howard Grotch, and Valery~A. Shelyuto.
\newblock {\em {Theory of Light Hydrogenic Bound States}}, volume 222.
\newblock Springer-Verlag, Berlin, 2007.

\bibitem{zhou2022simulation}
Lu-Ping Zhou, Xiao-Jie Ni, Zaher Salman, Andreas Suter, Jing-Yu Tang, Vjeran
  Vrankovic, and Thomas Prokscha.
\newblock {Simulation studies for upgrading a high-intensity surface muon
  beamline at Paul Scherrer Institute}.
\newblock {\em Phys. Rev. Accel. Beams}, 25(5):051601, 2022.

\bibitem{antognini2021mucool}
Aldo Antognini and David Taqqu.
\newblock {muCool: muon cooling for high-brightness $\mu^+$ beams}.
\newblock {\em SciPost Phys. Proc.}, 5:030, 2021.

\bibitem{2021_HiMB2}
M.~Aiba et~al.
\newblock {Science Case for the new High-Intensity Muon Beams HIMB at PSI}.
\newblock 11 2021.

\end{thebibliography}

\end{document}